\documentclass[conference]{IEEEtran}
\IEEEoverridecommandlockouts
\usepackage{cite}
\usepackage{amsmath,amssymb,amsfonts}
\usepackage{algorithmic}
\usepackage{graphicx}
\usepackage{textcomp}
\usepackage{enumitem}
\usepackage{multirow}
\usepackage{hyperref}   

\usepackage{soul}
\usepackage[table,x11names]{xcolor}
\def\BibTeX{{\rm B\kern-.05em{\sc i\kern-.025em b}\kern-.08em
    T\kern-.1667em\lower.7ex\hbox{E}\kern-.125emX}}
\usepackage{tcolorbox}  
\begin{document}


\title{An Investigation into Inconsistency of Software Vulnerability Severity across Data Sources}

\author{\IEEEauthorblockN{%
    Roland Croft\IEEEauthorrefmark{1}\IEEEauthorrefmark{2}, 
    M. Ali Babar\IEEEauthorrefmark{1}\IEEEauthorrefmark{2},
    Li Li\IEEEauthorrefmark{3}%
    }%
    \IEEEauthorblockA{\IEEEauthorrefmark{1} School of Computer Science, University of Adelaide, Adelaide, Australia, \{roland.croft, ali.babar\}@adelaide.edu.au}%
    \IEEEauthorblockA{\IEEEauthorrefmark{2} Cyber Security Cooperative Research Centre, Australia}%
    \IEEEauthorblockA{\IEEEauthorrefmark{3} Faculty of Information Technology, Monash University, Melbourne, Australia, \{li.li\}@monash.edu}%
}


\maketitle

\begin{abstract}
Software Vulnerability (SV) severity assessment is a vital task for informing SV remediation and triage. Ranking of SV severity scores is often used to advise prioritization of patching efforts. However, severity assessment is a difficult and subjective manual task that relies on expertise, knowledge, and standardized reporting schemes. Consequently, different data sources that perform independent analysis may provide conflicting severity rankings. Inconsistency across these data sources affects the reliability of severity assessment data, and can consequently impact SV prioritization and fixing. In this study, we investigate severity ranking inconsistencies over the SV reporting lifecycle. Our analysis helps characterize the nature of this problem, identify correlated factors, and determine the impacts of inconsistency on downstream tasks. Our findings observe that SV severity often lacks consideration or is underestimated during initial reporting, and such SVs consequently receive lower prioritization. We identify six potential attributes that are correlated to this misjudgment, and show that inconsistency in severity reporting schemes can severely degrade the performance of downstream severity prediction by up to 77\%. Our findings help raise awareness of SV severity data inconsistencies and draw attention to this data quality problem. These insights can help developers better consider SV severity data sources, and improve the reliability of consequent SV prioritization. Furthermore, we encourage researchers to provide more attention to SV severity data selection. 
\end{abstract}


\section{Introduction}
\label{sec:introduction}

Unpatched Software Vulnerabilities (SVs) can cause devastating consequences to organizations, and hence swift and effective remediation is essential \cite{shahriar2012}. Consequently, effective disclosure of SVs is a sensitive task. Identified SVs are often reported by multiple data sources, such as bug reports, vendor advisories, or vulnerability databases, to allow for timely and widespread dissemination \cite{dong2019towards}. 

With the sheer number of bugs and vulnerabilities that software developers encounter in modern software systems, developers also require intelligent and informed remediation plans \cite{dissanayake2021software}. Hence, SV severity assessment data is vital information provided by SV reporting sources that allows for better prioritization of fixing and patching efforts~\cite{fruhwirth2009improving}. The severity of an SV is often influenced by exploitability and impact factors \cite{le2021survey}; an SV that is both highly exploitable and has significant impacts should be considered critically severe. Severity rankings provide a natural ordering of SVs that can be used for initial prioritization. 


However, as SV reporting data sources provide information at different stages of the reporting lifecycle, inferred independently by different analysts and scoring schemes, various reports may provide conflicting severity rankings for the same sets of SVs. Prior works have observed variations in severity scores for aggregated vulnerability databases \cite{johnson2016can, jiang2020approach}. Whilst this inconsistency is expected due to the differences of each respective data source, it is an inherent problem. These independent data sources are used for the same tasks; SV assessment and prioritization. Hence, inconsistent severity rankings add confusion and unreliability. For instance, users may be unsure of which patches to prioritize if multiple data sources provide conflicting severity information. A set of vulnerabilities would be prioritized differently depending on the severity data source chosen. 

Furthermore, researchers also rely on the reliability and correctness of SV data. Prior research efforts have used SV data to derive development insights \cite{weicomprehensive}, or to develop automatic assessment approaches \cite{tian2016unreliability}. Particularly, SV severity prediction has been a commonly explored task \cite{le2021survey}. With conflicting information present in the available data sources however, the reliability and validity of such research is uncertain. 

Whilst prior works have observed severity inconsistency \cite{jiang2020approach}, or investigated discrepancies amongst common scoring schemes \cite{johnson2016can}, we are the first to analyse severity inconsistency within the full SV reporting lifecycle. Such insights are essential towards understanding severity information quality. We categorize three main SV reporting data sources \cite{dong2019towards}: 1) bug reports, which document initial identification; 2) vendor advisories, which perform initial disclosure; and 3) vulnerability databases that perform wider dissemination. We conduct our study through a large-scale investigation of the Mozilla Firefox development history, and its respective SV reporting sources: Bugzilla, the Mozilla Security Advisory, and the National Vulnerability Database (NVD). Our main contributions are: 
\begin{itemize}
    \item We provide an empirical investigation into the nature of inconsistency for severity rankings produced by different SV assessment data sources. Our analysis yields insights into the prevalence, characteristics, and correlated factors of such inconsistency. To the best of our knowledge, we are the first to examine such inconsistency across the full SV reporting lifecycle. 
    \item We quantify the impacts that variations in these data sources can have on SV prioritization and prediction. 
\end{itemize}

Primarily, we observe there to be weak correlation in the produced severity orderings for the three data sources that we have analyzed. This arouses a need for better standardization of the assessment information provided in these data sources, to increase reliability of prioritization schemes derived from such information. Furthermore, we observe that developer expertise is one of the key influential factors that leads to inconsistent severity rankings during initial reporting. We hence promote better education and training for this vital task. Finally, we also investigate the impacts that these issues can have for researchers, and observe that the choice of data source can heavily degrade the performance of severity prediction. We consequently advise caution and consideration for researchers of data quality when performing data selection. 

The remainder of this paper is organized as follows. Section \ref{sec:background} presents the background knowledge and motivation for this study. Section \ref{sec:method} details our research methodology. We present our findings in Section \ref{sec:results}, and discuss the implications of these findings in Section \ref{sec:implications}. Section \ref{sec:threats} details the potential threats to validity of this research. Section \ref{sec:related_work} describes related work. Finally, we conclude this paper in Section \ref{sec:conclusion}. We have made our dataset and scripts publicly available as a reproduction package~\cite{reproduction_package}. 

\section{Background}
\label{sec:background}
\subsection{Software Vulnerability Reporting Practices}
\label{sec:reporting}

\begin{figure}[bt]
  \centering
  \includegraphics[width=0.9\columnwidth]{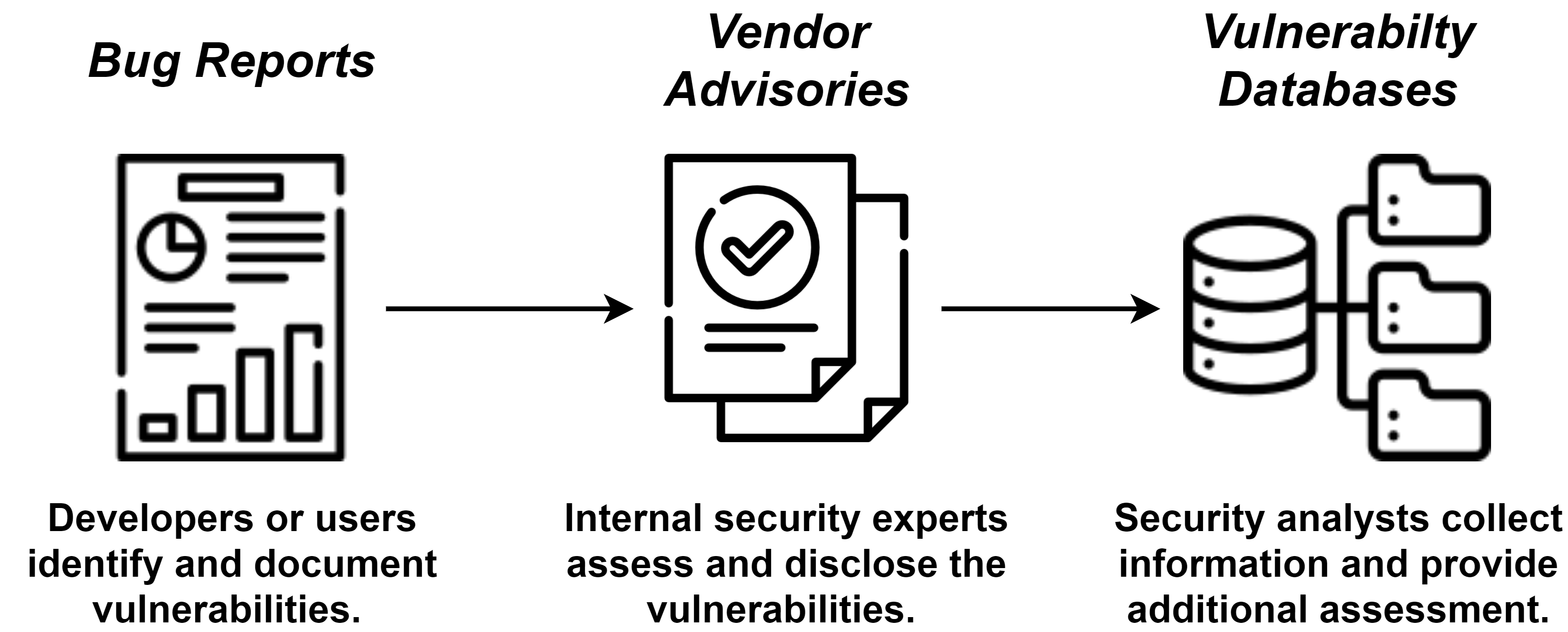}
  \caption{An overview of SV reporting processes.}
  \label{fig:reporting}
\end{figure}

As disclosure and dissemination of SVs is a sensitive task, both developers and security experts make efforts towards a thorough reporting lifecycle \cite{dong2019towards}. Figure \ref{fig:reporting} displays the standard SV disclosure process for large organisations, which contains three main stages: bug reports, vendor advisories, and vulnerability databases. Each stage provides independent severity assessment of SVs. We conducted our study on the Mozilla Firefox project, so we describe its respective data sources: Bugzilla\footnote{\url{https://bugzilla.mozilla.org/}}, the Mozilla Security Advisory\footnote{\url{https://www.mozilla.org/en-US/security/advisories/}}, and NVD\footnote{\url{https://nvd.nist.gov/}}. The flow of information in Figure \ref{fig:reporting} adds time delays in reporting for each data source. We found that SVs were disclosed in the Mozilla Advisory a median of 65 days after Bugzilla. Disclosure in NVD takes an additional median of 13 days. 

\textbf{Bug Reports.} Upon identification, SVs are reported to developers in the form of bug reports to a bug tracking system. Bug reports provide a description and basic assessment of identified bugs, to enable developers to implement a patch \cite{bettenburg2008makes}. Unlike the latter two data sources, bug report severity is not necessarily specific to SVs; regular defects are also assessed. However, bug report severity assessment is still highly important for SVs, as it is used to inform prioritization~\cite{bettenburg2008makes}.

Bugzilla is a popular bug tracking system that was originally developed and now used by the Mozilla project~\cite{Bugzilla}. Bugzilla reports are usually assigned a severity score by the reporter of the SV. There are four classes of severity that may be assigned\footnote{\url{https://wiki.mozilla.org/BMO/UserGuide/BugFields}}; \textit{S1-S4}. However, these ratings were only introduced in 2020, so we have manually mapped old Bugzilla severity classes to \textit{S1-S4} based on their definition: \textit{blocker (S1)}, the bug significantly impacts users or causes data loss; \textit{critical} or \textit{major (S2)}, the bug severely impairs functionality and a satisfactory workaround does not exist; \textit{normal (S3)}, the bug blocks non-critical functionality and a workaround exists; and \textit{minor (S4)}, the bug has low or no impact to users. 

\textbf{Vendor Advisories.} SVs that are recorded in a bug tracking system are also independently disclosed to the public through vendor advisories that describe and document all SVs of a particular vendor or product. The Common Vulnerabilities and Exposures (CVE) system provides a reference system for SVs \cite{CVE}. Vendor advisories often assign unique CVE IDs to SVs through a CVE Numbering Authority (CNA) \cite{CVE}. 

The Mozilla Foundation maintains the Mozilla Foundation Security Advisory \cite{mozillaadvisory}, and also acts as a CNA. This advisory is regularly updated, in correlation with new product releases. Severity ratings are independently assessed by the Mozilla Security team during the vulnerability remediation process \cite{mozillaadvisory}, and are also added as a keyword to the associated Bugzilla report once determined. The severity ratings are estimated through the expected exploitability and user impacts, and fall under four classes\footnote{\url{https://wiki.mozilla.org/Security_Severity_Ratings/Client}}: \textit{critical}, \textit{high}, \textit{moderate}, or \textit{low}. For example, High severity SVs are ``exploitable vulnerabilities which can lead to the widespread compromise of many users requiring no more than normal browsing actions.''

\textbf{Vulnerability Databases.} Lastly, vulnerability databases aggregate information from a variety of vendor advisories and other sources, to provide a standardized collection of disclosed SVs. Unlike the prior two data sources that are maintained internally by an organization, vulnerability databases are maintained externally by a third-party, and contain information relating to a variety of vendors and products. Besides NVD, there are many independently maintained vulnerability databases that operate towards similar purposes, such as Exploit-DB\footnote{\url{https://www.exploit-db.com/}}, Snyk Vulnerability Database\footnote{\url{https://snyk.io/vuln}}, or IBM X-Force\footnote{\url{https://exchange.xforce.ibmcloud.com/}}. 

However, vulnerability databases are not necessarily fully independent of each other as many aggregate information from a wide collection of resources, including other databases \cite{johnson2016can}. NVD is considered as the foremost vulnerability database due to its thorough maintenance and integration with CVE. Consequently, other databases exhibit heavy overlap with NVD \cite{johnson2016can}, so we only consider NVD for our analysis. 

NVD is built to synchronize with the CVE list, and hence waits for CVE IDs to appear. NVD analysts then examine each SV and add enhanced information \cite{NVD}, such as severity, type, and affected versions. NVD assigns a severity score to SVs using the Common Vulnerability Scoring System (CVSS) \cite{CVSS}. There are currently two active CVSS versions: CVSS 2 and CVSS 3. CVSS 3 was introduced in 2015 to account for the criticized lack of granularity of CVSS 2 \cite{CVSS}. However, we considered CVSS 2 for this study, as data for CVSS 3 is not as complete due to its late introduction, and previous SVs are still relevant in modern contexts. For CVSS 2, a severity ranking is assigned to SVs based on a numeric score that ranges from 0-10: \textit{Low} (0-3.9), \textit{Medium} (4.0-6.9), and \textit{High} (7.0-10.0). This score is calculated through a formula that assesses exploitability and impact metrics.

\subsection{Motivating Example}
As a motivating example, we consider the severity rankings and subsequent prioritization of a small subset of SVs. Table \ref{tab:sev_example} displays the severity rankings for each assessment source for all vulnerabilities fixed in Firefox 77 and disclosed in the Mozilla Advisory entry MFSA2020-20. Whilst each assessment source uses individual severity rating schemes, all schemes follow an ordinal scale. Although we outlined up to four levels of severity in Section \ref{sec:reporting}, each data source only uses three different categories for this example. Hence, we have converted each scheme to their relative rankings, with 1 being the most severe and 3 being the least severe. Table \ref{tab:pri_example} displays the natural ordering, and hence base prioritization scheme that would be inferred from these severity rankings. 

We can see that even for this small subset of vulnerabilities there are large variations in the rankings and subsequent ordering. For instance, only two out of the ten SVs (V2 \& V8) receive a common ranking across all three sources. Furthermore, V6 and V7 are prioritized first by the Mozilla Advisory, but last for NVD, despite both these data sources receiving assessment from security experts. 

This inconsistency adds unreliability and confusion to the severity data, and makes us question the accuracy of prioritization schemes inferred from this information. For instance, what values should we trust and which ordering is optimal? Furthermore, we question the validity of research outcomes derived from such datasets, as they would inherently change depending on the selected data source. These questions motivate our analysis into the full extent of this issue; its characteristics, factors, and impacts. 

\begin{table}[tb]
  \caption{Vulnerability rankings of Mozilla Firefox vulnerabilities that were fixed in release 77. 1 (red) is most severe, 2 (orange) is moderately severe, and 3 (yellow) is least severe.}
  \label{tab:sev_example}
  \centering
  \begin{tabular}{ccccc}
    \hline
    \multirow{2}{*}{\textbf{ID}} & \multirow{2}{*}{\textbf{Bugzilla ID}} & \multicolumn{3}{c}{\textbf{Severity Rankings}}\\
    \cline{3-5}
    & & \textbf{Bugzilla} & \textbf{Mozilla Advisory} & \textbf{NVD}\\
    \hline
    
    V1 & 1619305 & \cellcolor{orange}2 & \cellcolor{red}1 & \cellcolor{red}1 \\
    \hline
    V2 & 1620972 & \cellcolor{red}1 & \cellcolor{red}1 & \cellcolor{red}1 \\
    \hline
    V3 & 1623888 & \cellcolor{orange}2 & \cellcolor{yellow}3 & \cellcolor{orange}2 \\
    \hline
    V4 & 1625333 & \cellcolor{orange}2 & \cellcolor{red}1 & \cellcolor{red}1 \\
    \hline
    V5 & 1629506 & \cellcolor{orange}2 & \cellcolor{yellow}3 & \cellcolor{orange}2 \\
    \hline
    V6 & 1631576 & \cellcolor{yellow}3 & \cellcolor{red}1 & \cellcolor{yellow}3 \\
    \hline
    V7 & 1631618 & \cellcolor{red}1 & \cellcolor{red}1 & \cellcolor{yellow}3 \\
    \hline
    V8 & 1632717 & \cellcolor{red}1 & \cellcolor{red}1 & \cellcolor{red}1 \\
    \hline
    V9 & 1637112 & \cellcolor{yellow}3 & \cellcolor{orange}2 & \cellcolor{yellow}3 \\
    \hline
    V10 & 1639590 & \cellcolor{orange}2 & \cellcolor{red}1 & \cellcolor{red}1 \\
    \hline

\end{tabular}
\end{table}

\begin{table}[tb]
  \caption{The prioritization scheme for SVs fixed in Firefox 77 that is inferred directly from the severity rankings in table \ref{tab:sev_example}.}
  \label{tab:pri_example}
  \resizebox{\columnwidth}{!}{%
  \begin{tabular}{|c|p{2.4cm}|p{3.4cm}|p{2.4cm}|}
    \hline
    \multirow{2}{*}{\textbf{Priority}} & \multicolumn{3}{c|}{\textbf{Data Source}}\\
    \cline{2-4}
    & \textbf{Bugzilla} & \textbf{Mozilla Advisory} & \textbf{NVD}\\
    \hline
    
    1st & V2, V7, V8 & V1, V2, V4, V6, V7, V8, V10 & V1, V2, V4, V8, V10 \\
    \hline
    2nd & V1, V3, V4 ,V5, V10 & V9 & V3, V5 \\
    \hline
    3rd & V6, V9 & V3, V5 & V6, V7, V9 \\
    \hline
\end{tabular}%
}
\end{table}

\section{Research Methodology}
\label{sec:method}

\subsection{Research Questions}
To analyse the characteristics, causes and impacts of inconsistency for SV severity rankings across data sources, we aim to address the following three Research Questions (RQs): 

\textbf{RQ1.} \textbf{What is the nature of inconsistency for severity reporting schemes?}
We first aim to illustrate and characterize the nature of this problem by detailing the inconsistency of SV severity data and reporting schemes for three different SV data sources of the Mozilla Firefox dataset. We seek to raise awareness of this issue, so that developers may better judge the reliability of SV prioritization, and so that researchers have more consideration of severity data quality and consistency. 

\textbf{RQ2.} \textbf{Which factors influence inconsistent rankings for SV severity assessment during early stages?}
Aside from inconsistent scoring schemes and experts, variations in severity rankings can also be introduced from inaccurate assessment \cite{fruhwirth2009improving}. Misjudgment is most likely to occur during early SV assessment, due to less consideration and available information \cite{tian2016unreliability}. By identifying potentially correlated factors of initial SV severity inconsistency, we can speculate causal factors of misjudgment. These insights can potentially assist developers or project managers to pinpoint attributes that can improve SV severity assignment, and hence prioritization. 

\textbf{RQ3.} \textbf{How does SV severity data source inconsistency affect downstream predictions?}
Other than impacts to SV prioritization, severity data inconsistency can also impact researchers who analyse severity data. One such task that has been regularly investigated is severity prediction \cite{le2021survey}. Researchers can use different data sources depending on the selected stage for prediction, e.g., for bug reports \cite{tian2016unreliability} or SV databases \cite{le2019automated}. Through this RQ, we display the impacts on prediction performance that dataset selection can have, and hence motivate researchers to properly consider this issue.

\subsection{Data Collection}
We selected the Mozilla Firefox project for analysis for the following reasons. Firstly, it is one of the largest open-source web browsers with over 200 million active users as of September 2021 \cite{mozillastats}. Secondly, it has over a decade of active and publicly available development history. Thirdly, it maintains thorough and public SV reporting practices \cite{mozillaadvisory}. Finally, it has been the subject of many prior SV research studies \cite{shin2010evaluating,morrison2018vulnerabilities}. We collected SV data and severity information from each of the data sources described in Section \ref{sec:reporting}: Bugzilla, the Mozilla Advisory, and NVD. 

To form our dataset, we first identified known SVs for Mozilla Firefox through the Mozilla Advisory \cite{mozillaadvisory}. We then scraped all the reported SVs from a 10 year time period of June 21st 2011 to June 1st 2021 (Firefox Release 5 to 89). Advisory data prior to this time period was not available. This provided us with a set of 1503 advisory entries. 

\begin{figure}[tb]
  \centering
  \includegraphics[width=\linewidth]{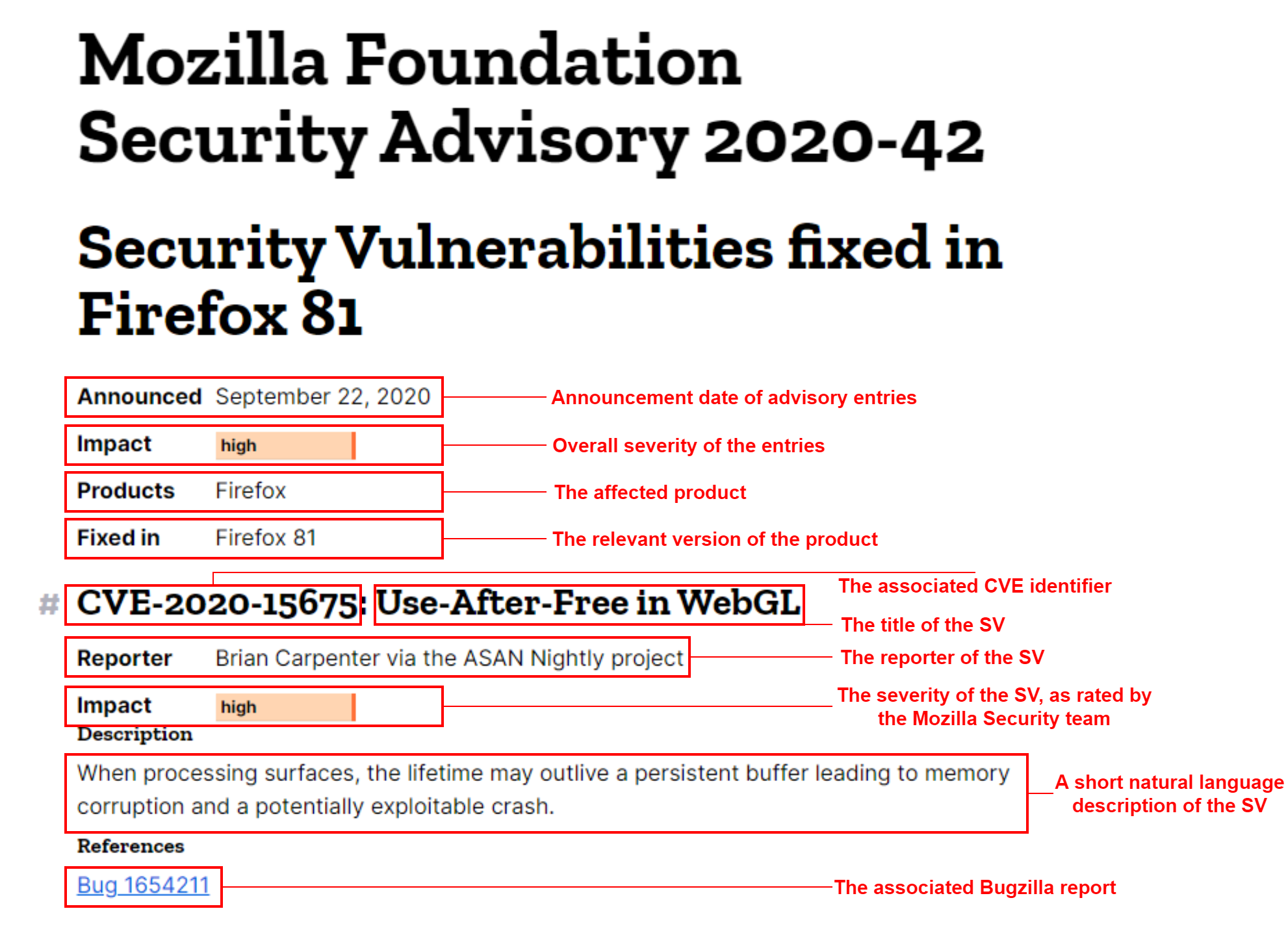}
  \caption{An annotated example of the Mozilla Foundation Security Advisory.}
  \label{fig:advisory_example}
\end{figure}

Figure \ref{fig:advisory_example} displays an annotated example of an advisory entry. Each advisory entry provides a link to an associated Bugzilla report and CVE ID. We collected data for each individual bug report, and split advisory entries that had multiple associated bug reports. Additionally, we scraped the NVD details of each entry, as identified through the CVE ID. A CVE ID may refer to multiple bug reports. We removed duplicate bug reports or bug reports that were private at the time of data collection. This provided us with a final set of 2455 unique SV bug reports for 1329 unique CVE IDs. 

\begin{table}[tb]
  \caption{Normalization of data source severity rankings.}
  \label{tab:normalization}
  \resizebox{\columnwidth}{!}{%
  \begin{tabular}{c|cccc}
    \hline
    \multirow{2}{*}{\textbf{Data Source}} & \multicolumn{4}{c}{\textbf{Severity Rankings}}\\
    \cline{2-5}
    & \textbf{1st} & \textbf{2nd} & \textbf{3rd} & \textbf{4th}\\
    \hline
    Bugzilla & blocker (S1) & critical/major (S2) & normal (S3) & minor (S4)\\
    \hline
    Mozilla Adv. & critical & high & moderate & low\\
    \hline
    NVD & critical & high & medium & low\\
    \hline
    
    \hline
    \textit{Normalized} & \textit{critical} & \textit{high} & \textit{medium} & \textit{low}\\
    \hline
\end{tabular}%
}
\end{table}

To enable comparison of the severity rankings from each source, we normalized the labels of each source to a common format, as displayed in Table \ref{tab:normalization}. For NVD, we added a \textit{Critical} severity class for CVSS 2 scores of 9.0-10.0. This allowed for better alignment of the other data sources, which had four classes each. We additionally added a \textit{none} severity class for entries that were missing severity information. 

We acknowledge that the severity classes of each source were not equivalent, due to differences in their definitions and scoring schemes. However, we considered their comparison valid as they used an ordinal ranking. We expect consistency in these rankings, as they are used to infer prioritization or other downstream tasks. 

\begin{table*}[tb]
  \caption{Attributes of a bug report that may influence SV severity misjudgment.}
  \label{tab:attributes}
  \resizebox{\textwidth}{!}{%
  \begin{tabular}{|p{1.5cm}|p{3.25cm}|p{6.5cm}|p{6cm}|}
    \hline
    \textbf{Category} & \textbf{Name} & \textbf{Definition} & \textbf{Rationale}\\
    \hline
    
    \multirow{3}{1.5cm}{Fix Difficulty} & Number of Patches & The number of patches that were submitted to fix the SV. & \multirow{3}{6cm}{If an SV is difficult to remediate, developers may not be able to properly assess it due to a lack of comprehension.}\\
    \cline{2-3}
    & Number of Files & The number of files edited across all patches. & \\
    \cline{2-3}
    & Number of Changes & The number of lines added, modified or deleted across all patches. & \\
    \hline
    
    \multirow{4}{1.5cm}{Review Process} & Fix Time & Time to fix a bug (resolved date - opened date). & Quickly resolved bugs may be ill-considered. \\
    \cline{2-4}
    & Number of Comments & The number of comments on the bug report. & \multirow{2}{6cm}{Extensive discussion may imply uncertainty of the evaluation of the SV.}\\
    \cline{2-3}
    & Number of Users & The number of users who comment on the bug report. & \\
    \cline{2-4}
    & Number of CC & The number of users in the mailing list for the bug report. & Heavy interest may imply better SV assessment. \\
    \hline
    
    \multirow{2}{1.5cm}{SV Nature} & Description Length & The number of words in the description. & More explanation may provide easier assessment.\\
    \cline{2-4}
    & Is Crash & Whether the string `crash' in the description. & Crashes have been shown to be indicative of severity label noise \cite{tian2016unreliability}. \\
    \hline
    
    \multirow{9}{2cm}{Reporter Expertise} & Reporter Bugs Filed & The number of bugs filed by the user. &  \multirow{2}{6cm}{Users who are good at reporting bugs may also be good at assessing them.}\\
    \cline{2-3}
    & Reporter Comments & The number of comments made by the user. & \\
    \cline{2-4}
    & Reporter Patches Submitted & The number of patches submitted by the user. & \multirow{6}{6cm}{Users who are good at resolving bugs may also be good at assessing them.}\\
    \cline{2-3}
    & Reporter Patches Reviewed & The number of patches reviewed by the user. & \\
    \cline{2-3}
    & Reporter Bugs Resolved & The number of bugs resolved by the user. & \\
    \cline{2-3}
    & Reporter Bugs Fixed & The number of bugs fixed by the user. & \\
    \cline{2-3}
    & Reporter Bugs Verified & The number of bugs verified by the user. & \\
    \cline{2-3}
    & Reporter Bugs Invalid & The number of bugs invalidated by the user. & \\
    \cline{2-4}
    & Reporter Profile Age & The profile age of the user (current date - creation date). & Older users may be better at assessing SVs. \\

    \hline
\end{tabular}%
}
\end{table*}

\subsection{Statistical Modeling}
To address RQ2, we aimed to identify factors that may lead to inconsistency in severity rankings for the initial Bugzilla severity assessment. As Bugzilla reports received the least security consideration, due to their fast communication, we speculate that inconsistencies to the Mozilla Advisory can be associated with potential misjudgment. We assume that correct Bugzilla severity assessment would induce severity rankings consistent with ones assigned and evaluated by experts. 

Hence, to identify correlated factors for SV severity misjudgment, we used statistical tests and regression analysis, following practices established by prior works \cite{mcintosh2016empirical,paul2021security}. Our dependent variable was whether Bugzilla severity was consistent with Mozilla Advisory severity. We did not consider NVD for this RQ as it is not as closely linked to Bugzilla reports as the Mozilla Advisory, and hence may introduce additional confounding factors. For explanatory variables, we scraped 18 attributes describing bug report metadata, as inspired by previous works \cite{mcintosh2016empirical,paul2021security}. Table \ref{tab:attributes} displays the 18 attributes and our rationale behind their inclusion for our task. 

\begin{figure}[tb]
  \centering
  \includegraphics[width=0.85\linewidth]{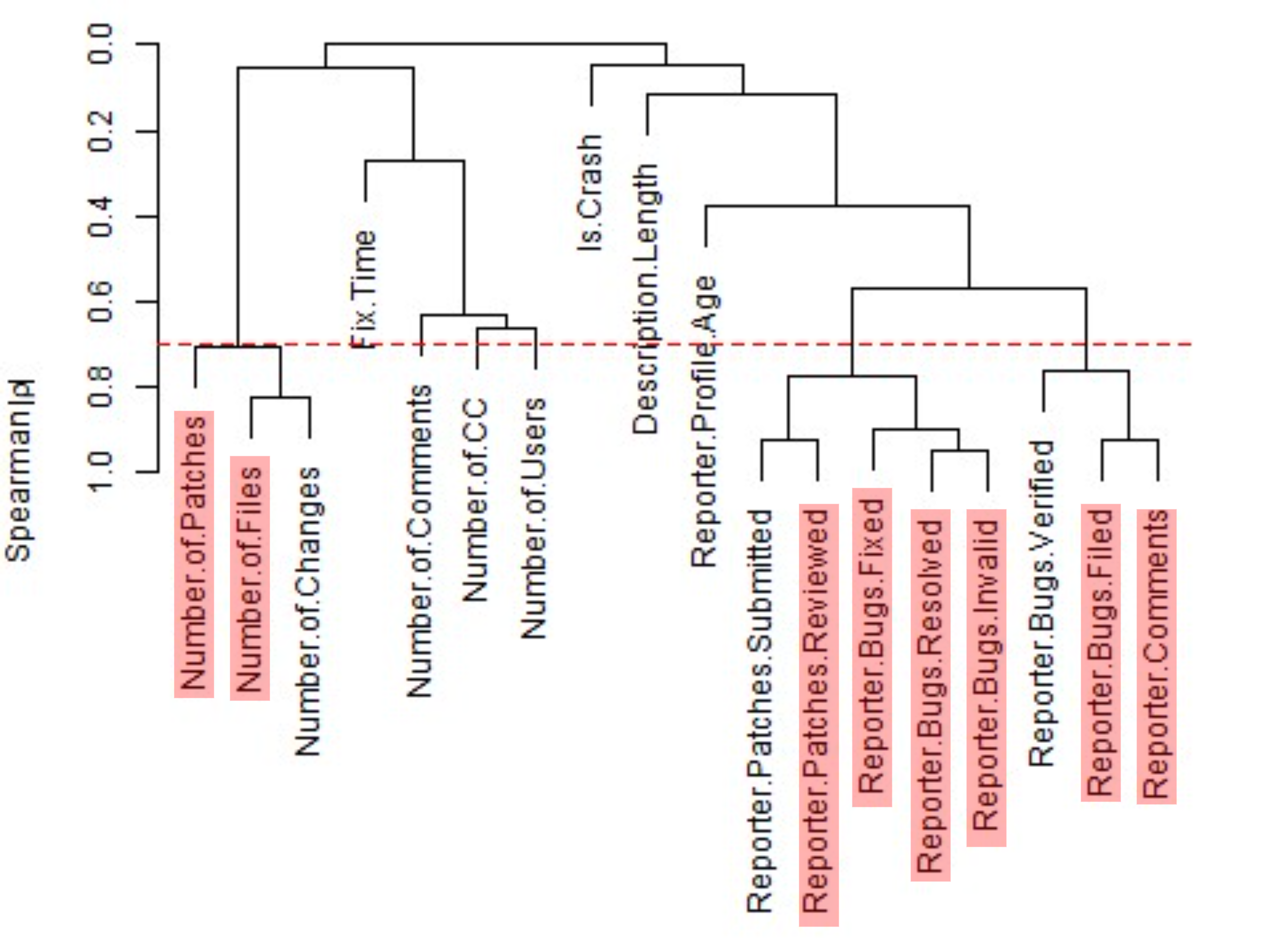}
  \caption{Hierarchical clustering of variables according to Spearman's rank-order correlation. The dashed line indicates the high correlation threshold of 0.7. Removed variables are highlighted in red.}
  \label{fig:correlation}
\end{figure}

We first removed correlated attributes by using Spearman's rank-order correlation test \cite{spearman1987proof} to determine highly correlated explanatory variables. We only retained one variable from each grouping that falls above a threshold value of 0.7, as this threshold value has been recommended in prior SE studies \cite{mcintosh2016empirical,paul2021security}. Figure \ref{fig:correlation} displays the output of this correlation analysis; eight variables were removed. Attributes within the Fix Difficulty and Reporter Expertise categories were highly correlated to each other. 

We then fitted a logistic regression model to the remaining 10 explanatory variables, as logistic regression models are simple but effective predictors for binary response variables \cite{wright1995logistic}. We used the z-value of the regression coefficient to determine which variables had statistically significant coefficients, and were hence correlated with consistent SV severity assessment. 

For RQ3, we aimed to investigate the impacts that data source can have on bug or SV severity prediction; a common problem explored by researchers in prior literature. Hence, we replicated standard practices by using textual description data as input \cite{tian2016unreliability,le2019automated}, to predict the normalized severity categories described in Table \ref{tab:normalization}. Following these practices, we preprocessed text descriptions through removal of stop words (using the NLTK and sklearn stopword list) and punctuation, conversion to lowercase, and stemming. The descriptions were then encoded using a bag-of-words model; we only extracted features for words that appeared in more than 0.1\% of all descriptions \cite{le2019automated}. We evaluated the same classifiers and tuned the same hyperparameters that were experimented with by prior works \cite{le2019automated}. Full details of the experimental setup are available in our reproduction package \cite{reproduction_package}. 

Time-based validation methods have been shown to be important for this line of research \cite{le2019automated}. Hence, we sorted our dataset by submission date of the report description, and then divided the dataset into a training, validation and test set through an 80:10:10 split. The validation set was used for tuning hyperparameters and selecting a model, and the test set was used for final evaluation of the optimal model. Model performance was measured using Matthew's Correlation Coefficient (MCC), as it has been found to be a less biased metric than other evaluation metrics \cite{yao2020assessing}. The MCC score can range from -1 to 1, where 1 is the best value. 

To thoroughly investigate the inconsistency of the SV severity labels across these data sources, we evaluated prediction performance for each SV description dataset when using each set of severity labels. For example, we built three prediction models for NVD descriptions when using NVD, Bugzilla and Mozilla Advisory labels separately. 

\section{Results}
\label{sec:results}

\begin{figure*}[tb]
  \centering
  \includegraphics[width=0.85\textwidth]{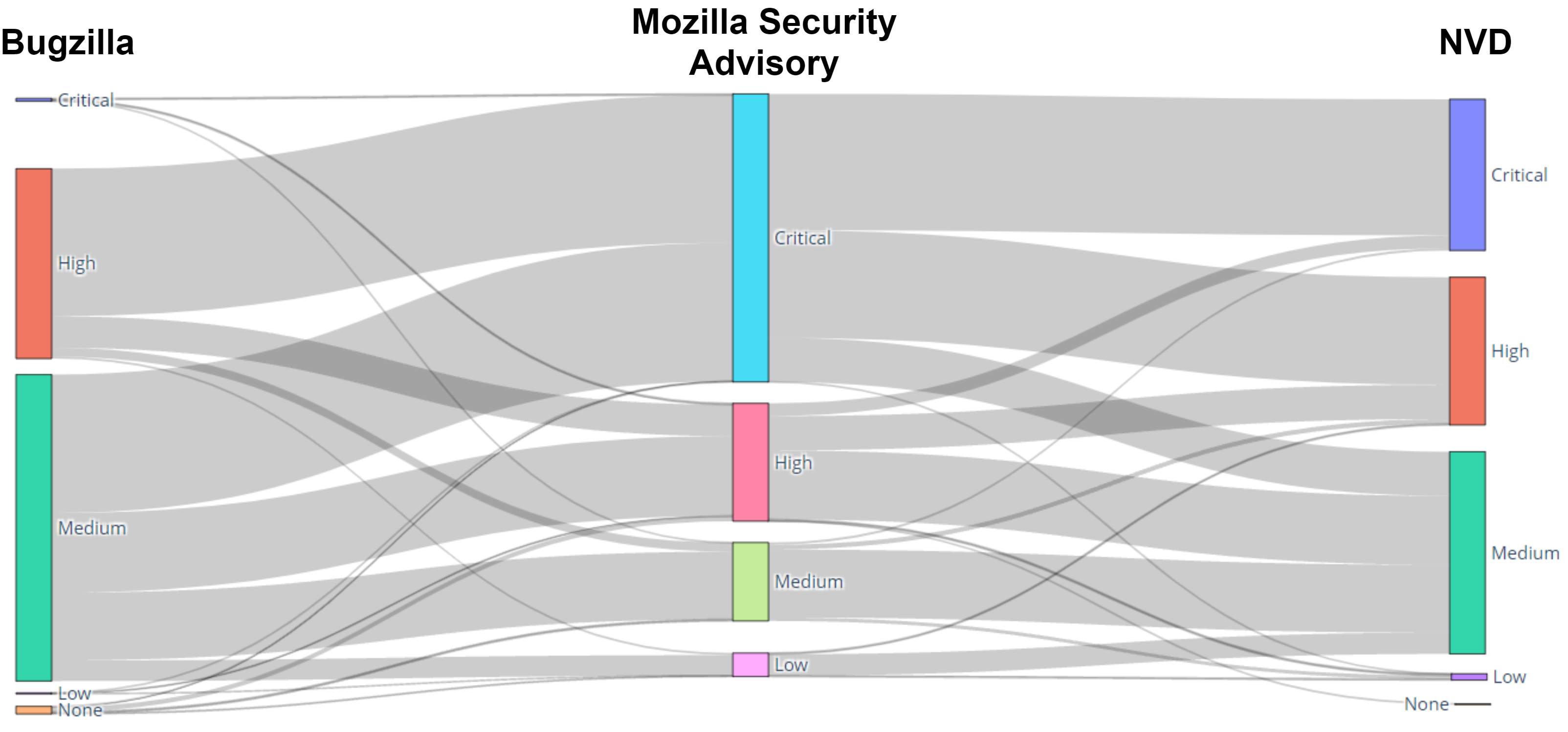}
  \caption{Change in severity rankings across data sources.}
  \label{fig:distribution}
\end{figure*}

\subsection{RQ1. What is the nature of inconsistency for severity reporting schemes?}
\begin{figure}[tb]
  \centering
  \includegraphics[width=\columnwidth]{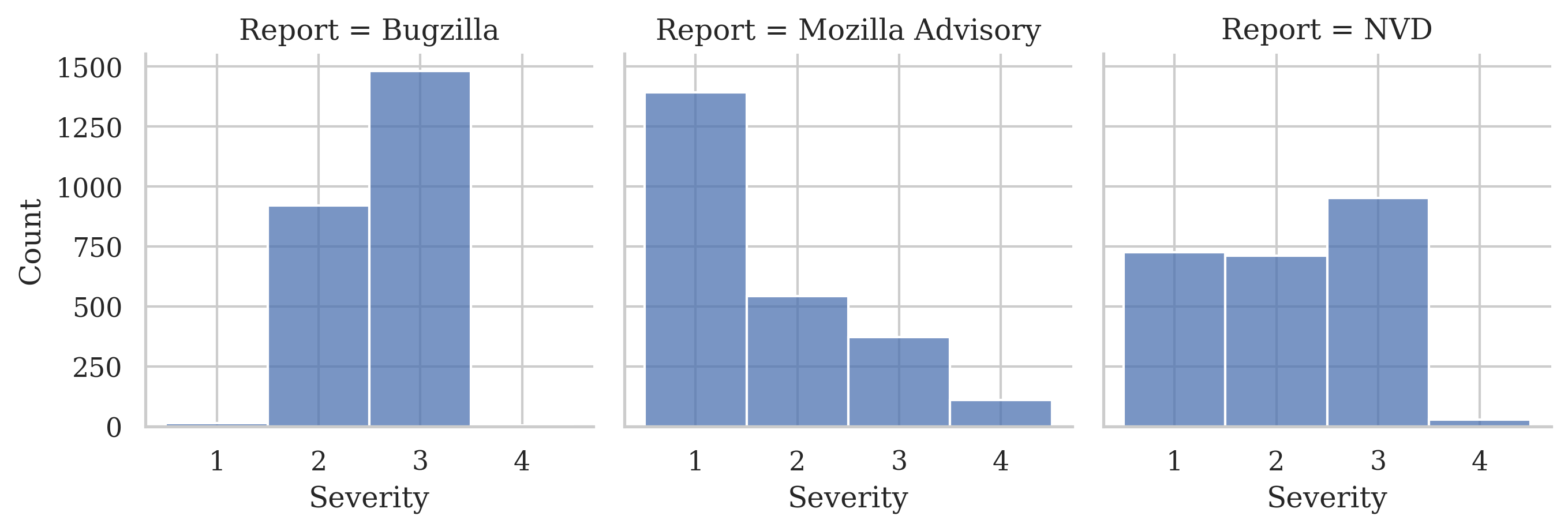}
  \caption{Severity ranking distribution for each data source, where 1 = critical severity and 4 = low severity.}
  \label{fig:histplot}
\end{figure}

Figure \ref{fig:distribution} displays the changes in severity rankings across the severity reporting schemes of the three data sources. Figure \ref{fig:histplot} presents a histogram of severity rankings for each data source. 

Bugzilla severity rankings were more conservative; the majority of SVs were documented as Medium severity, and very few SVs were documented as Low or Critical severity. This limited diversity for Bugzilla reports may limit their practical usefulness, as most SVs would receive the same ordering under the Bugzilla severity scheme. Unlike Bugzilla, the Mozilla Security Advisory was skewed towards the most severe ranking. NVD exhibited the most uniform range of severity rankings, but still contained very few entries of Low severity. The flow of severity rankings displayed in Figure \ref{fig:distribution} reflected these changes in distribution. 

SV severity is predominantly underestimated using the Bugzilla reporting scheme; 74\% (1807/2455) and 48\% (1169/2455) of Bugzilla reports were assigned severity rankings lower than that of the Mozilla Advisory and NVD, respectively. This implies that Bugzilla reports either pay little consideration to severity rankings, or lack the proper information and expertise to correctly assess SVs at reporting time. However, this incorrect assessment can lead to dire consequences, as this initial reporting information is used to establish prioritization.  We found that over 94\% (2317/2455) of bug reports were assigned a priority score using the initial information and severity data, before any security assessment from the Mozilla security team had occurred. Using a Mann-Whitney U test \cite{mann1947test}, we were able to favor the alternate hypothesis that Bugzilla SVs were often assigned a lower priority score when their severity rankings were inconsistent with the Mozilla Advisory ($U=410981, p<0.001$). This lower prioritization due to improper SV assessment can lead to delays in remediation. 

\begin{tcolorbox}[right=1pt, left=1pt, top=1pt, bottom=1pt, colback=white]
    \textbf{Finding 1:} \textit{SV severity is often underestimated during first reporting. Misjudged SVs often receive lower priority.}
\end{tcolorbox}


Furthermore, using Spearman's rank-order correlation coefficient \cite{spearman1987proof}, we observed Bugzilla severity rankings exhibited a weak correlation to the rankings of both the Mozilla Advisory ($\rho=0.346, p<0.001$) and NVD ($\rho=0.260, p<0.001$). This may be expected, as the Bugzilla severity scoring scheme is not exclusive to SVs, unlike the other two data sources. However, Bugzilla severity scores are still important as they are used to infer bug report prioritization. Hence, the lack of consistency between rankings for these data sources suggests that initial prioritization inferred from Bugzilla scoring schemes may be ill-suited for these security critical bugs. 

There were also considerable differences in the severity rankings between the Mozilla Advisory and NVD, despite both of these sources receiving expert security analysis. Over 44\% (1092/2455) of SVs received a higher severity ranking in the Mozilla Security Advisory than they did on NVD. Whilst it is unclear which severity ratings are more correct, this inconsistency at least implies that users ought to be aware of the nature and characteristics of each vulnerability database. 

\begin{tcolorbox}[right=1pt, left=1pt, top=1pt, bottom=1pt, colback=white]
    \textbf{Finding 2:} \textit{Security experts and databases report SV severity differently.}
\end{tcolorbox}

However, the Mozilla advisory and NVD severity rankings had a moderate to strong Spearman's rank-order correlation coefficient ($\rho=0.624, p<0.001$). Whilst their consistency was far from perfect, these two expert-based security sources at least had more correlation than the Bugzilla scoring scheme. This implies that expert assigned severity scores are more reliable than initial assessment scores. 

We further observed that NVD severity rankings were not noise free either. Eight of the SVs in our dataset were referenced by two or more CVE IDs. This typically arose when SVs were announced in batches in the Mozilla Advisory, and were individually unspecified. Despite this duplication, the assigned severity was not always consistent. Five out of the eight (63\%) duplicate NVD entries had different CVSS 2 scores. For instance, CVE-2016-1953 and CVE-2016-2805 were reported from a common bug report, but the former was assigned medium severity whereas the latter was assigned critical severity. 

\begin{tcolorbox}[right=1pt, left=1pt, top=1pt, bottom=1pt, colback=white]
    \textbf{Finding 3:} \textit{Expert maintained vulnerability databases can still contain inconsistencies.}
\end{tcolorbox}

\subsection{RQ2. Which factors influence inconsistent rankings for SV severity assessment during early stages?}

\begin{table}[tb]
  \centering
  \caption{Statistical significance of the coefficients of bug report attributes for predicting whether the SV severity ranking will be correctly assessed.}
  \label{tab:coefficients}
  \begin{tabular}{ccc}
    \hline
    \textbf{Variable} & \textbf{\textit{$z$}} & \textbf{\textit{$P>|z|$}}\\
    \hline
    \rowcolor{red} Reporter Patches Submitted & -2.966 & 0.003\\
    \rowcolor{green} Fix Time & 4.821 & \textless 0.001\\
    \rowcolor{red} Reporter Profile Age & -7.801 & \textless 0.001\\
    Number of CC & -0.092 & 0.927\\
    \rowcolor{red} Description Length & -5.016 & \textless 0.001\\
    Number of Comments & -1.088 & 0.277\\
    Number of Users & 0.599 & 0.549\\
    Number of Changes & -0.769 & 0.442\\
    \rowcolor{red} Is Crash & -4.467 & \textless 0.001\\
    \rowcolor{red} Reporter Bugs Verified & -2.639 & 0.008\\
    \hline
\end{tabular}
\end{table}

Aside from independent severity scoring schemes, misjudgment is another factor that can lead to inconsistency \cite{fruhwirth2009improving}. Inaccurate assessment or lack of consideration by users of a particular data source will add variation to severity rankings. For RQ2, we investigated potential causal factors that may lead to inconsistency between Bugzilla and Mozilla Advisory severity rankings, and hence potential misjudgment. We make the assumption that misjudgment is most likely to occur during early stages of the reporting lifecycle, as these stages receive the least consideration. Bugzilla severity scores are not as critically examined as the Mozilla Advisory ratings, which undergo thorough review for public disclosure. 

We applied regression analysis to identify quantitative variables that were correlated to consistent severity assessment for Bugzilla reports. Table \ref{tab:coefficients} displays the explanatory power for each of the attributes for predicting the likelihood that SV severity will be misjudged. We considered statistically significant coefficients as those with a p-value \textless 0.01. Positively correlated significant coefficients are highlighted in green, whereas negatively correlated significant coefficients are highlighted in red. A positive correlation implies that an increase in the value of this attribute will increase the likelihood that SV severity is correctly assigned, whereas a negative correlation implies that an increase in this value will increase the likelihood that SV severity is misjudged. 

We evaluated the goodness of fit of our logistic regression model by calculating the Area Under the Curve of the Receiver Operating Characteristic (AUC-ROC), as similarly used in prior works \cite{mcintosh2016empirical,paul2021security}. Our model achieved an AUC-ROC value of 0.687. Although this is not a particularly strong value, we deemed it to be sufficient as regression coefficients are unaffected by goodness of fit \cite{d1986goodness}.

Interestingly, the reporter expertise variables from Table \ref{tab:attributes} were negatively correlated with correct severity assessment. Users with more experience, both through profile age and practical experience in submitting patches and verifying bugs, generally underestimated the severity when reporting an SV. This is unexpected, as one would often assume that users with more expertise would be better equipped to assess severity. Whilst we are unable to provide an explicit reason for this phenomenon, we can provide some speculation. Firstly, more experienced users may be busier or become less cautious, and hence have less time to spend filling reports. Furthermore, although these developers may be experienced, they still may not have the necessary security knowledge to accurately assess these unique defects, and triaging of security defects may be substantially different. Over 97\% of the Bugzilla severity scores were assigned by the initial bug reporter. 

\begin{tcolorbox}[right=1pt, left=1pt, top=1pt, bottom=1pt, colback=white]
    \textbf{Finding 4:} \textit{Reporters with more experience are more likely to incorrectly assess SV severity.}
\end{tcolorbox}

Bug reports with longer descriptions also had an increased likelihood of misjudgment. This is again unexpected as we would suppose that the more information that is available, then the more informed the SV severity assessment would be. However, upon inspection we found long Bugzilla descriptions to typically come from lengthy crash/error logs and stack traces. SVs relating to crashes were also generally underestimated. This is perhaps because they were so common; 880 of the 2455 (36\%) SVs had associated crash data.

\begin{tcolorbox}[right=1pt, left=1pt, top=1pt, bottom=1pt, colback=white]
    \textbf{Finding 5:} \textit{Bug reports related to crashes or containing lengthy descriptions are more likely to be misjudged.}
\end{tcolorbox}

Inversely, SVs that took a long time to fix were more likely to be assessed correctly. This may be because reporters who can recognize SVs that require more effort to fix also expend more effort towards assessing them. 

\begin{table}[tb]
  \centering
  \caption{Number and percentage of consistently assessed SVs for the five most dominant CWE categories (cases $>$ 30).}
  \label{tab:cwe_analysis}
  \begin{tabular}{p{1cm}p{3cm}p{1cm}p{1cm}p{1cm}}
    \hline
    \textbf{CWE ID} & \textbf{Category} & \textbf{\# Cases} & \textbf{\# Consistent} & \textbf{\% Consistent}\\
    \hline
    CWE 664 & Improper Control of a Resource Through its Lifetime & 1566 & 239 & 15.26\\
    \hline
    CWE 707 & Improper Neutralization & 188 & 91 & 48.4\\
    \hline
    CWE 264 & Permissions, Privileges \& Access Controls & 123 & 44 & 35.77\\
    \hline
    CWE 693 & Protection Mechanism Failure & 34 & 14 & 41.18\\
    \hline
    CWE 399 & Resource Management Errors & 32 & 5 & 15.63\\
    \hline
\end{tabular}
\end{table}

Additionally, we examined whether SV type affects severity assessment. Different SV types exhibit different characteristics, hence one would expect that difficulty of assessment for each SV type also varies. The Common Weakness Enumerations (CWE) catalogs over 1000 different categories of SVs, using a hierarchical structure. We grouped CWEs to their highest level category to lower the dimensionality of our analysis, similar to Croft et al. \cite{croft2021empirical}. If a CWE is contained by more than one category, we assigned it to the most frequent one. To increase the validity of our statistical analysis, we only considered high-level CWE categories that occurred more than 30 times in our dataset. This excluded 512 bug reports from the statistical test. 

Using a Chi-Square test of independence \cite{pearson1900x}, it was statistically significant that some CWE types were harder to assess than others ($X^2=153.74, p<0.001$). Table \ref{tab:cwe_analysis} displays the number and proportion of SV types consistently (and inconsistently) assessed for the most frequent CWE categories. 

As discussed in RQ1, Bugzilla severity rankings were generally inconsistent with the severity rankings assessed by the Mozilla security team. From Table \ref{tab:cwe_analysis} it can be seen that resource related SV types (CWE 664 and CWE 399) were particularly difficult for reporters to correctly assess in Bugzilla reports; approximately 85\% of severity rankings for these SVs were misjudged. Resource control and management can relate to a variety of weaknesses, such as crashes, race conditions, and memory leakage. These SVs may be difficult to assess due to their high frequency; reporters may not consider them as severe if they occur frequently. Furthermore, resource related SVs do not have as a distinct security context as other SVs, which may lead reporters to underestimate their severity. Improper Neutralization (CWE 707) and Access Control related SV types (CWE 264 and CWE 694) were more frequently consistent with the Mozilla Advisory severity rankings. This may be because the security impacts of these types are more well known \cite{barnum2005knowledge} and hence easier to assess. 

\begin{tcolorbox}[right=1pt, left=1pt, top=1pt, bottom=1pt, colback=white]
    \textbf{Finding 6:} \textit{SV type influences severity assessment accuracy. Resource related SVs are harder to assess.}
\end{tcolorbox}

\subsection{RQ3. How does SV severity data source inconsistency affect downstream predictions?}
Table \ref{tab:prediction_results} displays the performance of the tuned prediction models for each data source. The performance for all tasks was lower than the reported performance of prior works \cite{tian2016unreliability,le2019automated}. We attribute this to our smaller dataset size in comparison to prior works, as otherwise our implemented techniques were largely similar. We were able to reproduce the performance reported by Le et al. \cite{le2019automated} when using all documented NVD entries, not just for Mozilla Firefox. 

\begin{table}[tb]
  \centering
  \caption{Severity prediction model performance for different SV description and severity ranking data sources. \% Change indicates the percentage change in MCC from using a uniform description and severity data source.}
  \label{tab:prediction_results}
  \begin{tabular}{cccc}
    \hline
    \textbf{Description Source} & \textbf{Severity Source} & \textbf{MCC} & \textbf{\% Change}\\
    \hline
    \hline
    Bugzilla & Bugzilla & 0.184 & -\\
    \hline
    Bugzilla & Mozilla Advisory & 0.042 & -77.17\\
    \hline
    Bugzilla & NVD & 0.124 & -32.61\\
    \hline
    \hline
    Mozilla Advisory & Mozilla Advisory & 0.16 & -\\
    \hline
    Mozilla Advisory & Bugzilla & 0.051 & -68.13\\
    \hline
    Mozilla Advisory & NVD & 0.231 & +44.38\\
    \hline
    \hline
    NVD & NVD & 0.217 & -\\
    \hline
    NVD & Bugzilla & 0.062 & -71.43\\
    \hline
    NVD & Mozilla Advisory & 0.086 & -60.37\\
    \hline
\end{tabular}
\end{table}

We observe the performance to differ for different data sources. When using a uniform description and severity label source, the NVD data source produced the best performing models. This high performance is likely credited to the targeted analysis and standardized reporting of the NVD; both the descriptions and severity rankings follow consistent guidelines and are reviewed by security experts. Bugzilla data performed worse than NVD, likely as it was missing consistency. Descriptions and severity rankings were often unchecked and written by various reporters. One may expect Bugzilla to perform better, as the Bugzilla descriptions were usually longer than the other sources and hence may have provided more information. However, as observed in \textit{Finding 5}, lengthy descriptions often contained noisy crash/error logs, which appear to be uninformative to the prediction model. The imbalanced distribution of severity ranking classes for Bugzilla data may also negatively influence the prediction performance. Finally, the Mozilla Security Advisory data performed worst, potentially because many advisory entries were missing meaningful descriptions. Entries that contained a batch of bug reports were not provided specific descriptions, e.g., the last two entries of MFSA2021-23\footnote{\url{https://www.mozilla.org/en-US/security/advisories/mfsa2021-23/\#CVE-2021-29967}}. The Mozilla Advisory appears to aim for simplicity, but this makes it a poor information source for data-driven tasks. 

This creates an interesting decision for researchers, as each of these data sources has a different time gap for producing its description after the initial SV detection. Whilst NVD performs the best, it has the largest average time delay. Bugzilla produces descriptions instantaneously, but does not perform as well. 

\begin{tcolorbox}[right=1pt, left=1pt, top=1pt, bottom=1pt, colback=white]
    \textbf{Finding 7:} \textit{Different data sources provide different performance for SV severity prediction.}
\end{tcolorbox}

We also investigated transferring labels to descriptions of other data sources. We aim to investigate the extent of disparity between these sources. If any similarities exist between them, then knowledge should be able to be transferred from one source to another. For instance, can bug report descriptions also serve as a predictor of the severity labels assigned by security experts, or vice versa. However, the performance for the majority of these models noticeably dropped under this scenario; performance decreased by 77\% in the worst case, which further highlights inconsistency in severity rankings. 

The performance most noticeably degraded when using NVD descriptions with other severity label sources. If we consider NVD labels to be the most correct, as they are assessed at the latest stage and maintained by various security experts, then this performance decrease is expected as we are adding noise and inaccuracies to the labels by changing data source. The performance decreased by 71\% when using Bugzilla labels in comparison to NVD labels. This again may be due to the heavy class imbalance of Bugzilla rankings. Conversely, model performance actually increased for the Mozilla Security Advisory descriptions when using NVD severity data, which attests to the quality of these labels. However, performance still decreased when Bugzilla reports were assigned NVD severity data. This may imply that Bugzilla descriptions have insufficient information to assess severity. 

\begin{tcolorbox}[right=1pt, left=1pt, top=1pt, bottom=1pt, colback=white]
    \textbf{Finding 8:} \textit{Using inappropriate severity data sources (i.e. inconsistent labels) degrades prediction performance.}
\end{tcolorbox}

\section{Implications}
\label{sec:implications}


\subsection{For Developers and Managers}
We first encourage SV reporters to exhibit more caution and security consideration when assessing SVs. Through \textit{Finding 1}, we observed that developers often underestimated the severity of SVs during initial reporting in Bugzilla. Whilst this may be expected, due to the lack of dedicated and delayed analysis like the other two data sources, proper SV assessment is still vital at this stage. We found that 94\% of bug reports were prioritized purely based on the Bugzilla severity ranking, and misjudgment often led to lower prioritization. By raising awareness of this issue, we hope our findings can motivate developers to spend more time and efforts on initial severity assessment. Managers may also choose to implement security specific assessment schemes for bug reports. 

Accurate SV assessment can be difficult however. This is further observed through \textit{Findings 4} and \textit{6}, which suggest that inconsistencies in early stage assessment rankings are caused by a lack of developer expertise or knowledge. We promote the need for developers to be given adequate security training to perform this assessment. We found that assessment rankings were particularly inconsistent for Resource Management Errors (CWE-399) and Improper Control of a Resource Through its Lifetime (CWE-664). Hence, education should especially be provided for these types, so that developers can better recognize and assess them. 

Furthermore, we suggest developers be wary of prioritization schemes directly inferred from base severity orderings. The inconsistency of severity rankings that we have observed in \textit{Findings 1} and \textit{2} implies that prioritization schemes that are solely dependent on this information are unreliable. If additional assessment efforts and resources are available, developers may find benefit from more intelligent prioritization schemes that also consider appropriate context, such as vulnerability conditions and exploit maturity, rather than just theoretical assessment. CVSS offers the ability for this extra assessment, through additional \textit{temporal} and \textit{environmental} metrics that can be assigned on top of the base score.

\subsection{For Researchers}
Firstly, researchers ought to continue to provide tool and knowledge support for SV severity assessment. Much research has been conducted into automated severity prediction \cite{gomes2019bug}, but this work is yet to make inroads into software development practices. Through \textit{Findings 1, 4} and \textit{6}, we observe that support is particularly needed for developers during early stage assessment of bug reports. 

Without existing standardization, we similarly encourage researchers to investigate more varied severity data sources. Whilst much research has been done regarding severity prediction for NVD, very little work has been performed on SV severity prediction for bug reports or vendor advisories \cite{le2021survey}. Existing bug report severity research only focuses on regular software defects \cite{gomes2019bug}. However, in \textit{Findings 2} and \textit{6} we see that there is a heavy variation amongst these data sources. This may pose threats to external validity of existing research, as findings derived from one data source may not generalize to others. Hence, developers ought to consider more data sources due to their inconsistencies, or develop more robust methods that can handle such variation. 

Whilst we encourage researchers to use more varied data sources, we also urge researchers to have more consideration for the impacts of their data selection. From \textit{Findings 7} and \textit{8}, we observed that the choice and consistency of data sources had significant impacts on severity prediction. Hence, researchers should be aware of noise and quality for their selected datasets. 

Finally, we also suggest researchers to provide support for project specific severity assessment. We speculate that consideration of environmental and temporal metrics, such as those optionally available in CVSS, may provide more reliable and consistent assessment data. However, developers need support in acquiring this assessment, as current data sources only consider generalized base metrics. 

\subsection{For Security Experts}

There is a need for standardization of SV severity assessment by security experts. A current lack of standardization results in different data sources producing different severity rankings and hence prioritization schemes, as seen in \textit{Finding 2}. This inconsistency adds confusion and a lack of reliability to this data. Furthermore, a lack of standardized data sources negatively impacts the external validity of derived research outcomes, as previously discussed. 

Whilst adoption and use of the Common Vulnerability Scoring Scheme (CVSS) \cite{CVSS} has made great strides towards standardized severity assessment, this issue is still far from solved. Independent analysis of SVs has been demonstrated to still lead to varying CVSS scores \cite{johnson2016can, jiang2020approach, schweigler2020investigation}, due to potentially subjective and inaccurate assessment. Furthermore, use of the CVSS is also not commonplace yet; it was only utilized by one of our three data sources. Hence, there is a need for improved solutions towards standardization of severity assessment. Some solutions have been suggested for this problem, such as meta-scores \cite{VulDB} or majority voting schemes \cite{jiang2020approach}. However, the validity and effectiveness of these solutions are currently un-evaluated. 

Furthermore, we encourage security experts to make more efforts towards cleaning the records within these data sources, as this is another source of inconsistency. In \textit{Finding 3} we observed some inconsistency within NVD from duplicate references to bug reports, and other researchers have observed similar noise issues \cite{anwar2021cleaning}. It is the responsibility of security experts who maintain these data sources, to also ensure the quality of their documented data. Whilst security experts already make substantial efforts towards maintenance and accuracy, more thorough review and quality control of these data sources will potentially be a valuable effort.

\section{Threats to Validity}
\label{sec:threats}

\textit{Construct Validity:} A potential threat is our mechanism of comparison for SV severity across data sources. Due to the independence of these data sources, inconsistency is expected and may arise from a variety of factors. The ranking produced by one source would not be directly equivalent to that of another. Additionally, some manipulation of the categorization schemes was required to allow for comparison. The use of CVSS 2 instead of CVSS 3 can also lead to some potential information loss from the assessment. However, we believe comparison to still be valid due to the ordinal scale of each scheme, and assert that inconsistency is an issue despite data source independence. 

\textit{Internal Validity:} We acknowledge that there may be confounding factors influencing correlation for RQ2. For instance, the reporter ID or the affected software product may also have high correlation to SV severity misjudgment. We will investigate these potential factors in future work. 

\textit{External Validity:} Like most empirical studies, our results may not sufficiently generalize to other applications or datasets. However, we have conducted our analysis on the Mozilla Firefox dataset, which is a large open-source application with many users. Furthermore, this dataset has been commonly analysed in prior research \cite{shin2010evaluating,morrison2018vulnerabilities}. 

\textit{Conclusion Validity:} As we heavily relied on statistical analysis and hypothesis testing to infer our findings, our study may suffer from conclusion validity. However, we believe that the dataset we have used is sufficiently large to draw conclusions from. Furthermore, we have used a strict threshold of \textit{p \textless 0.01} for null hypothesis rejection.

\section{Related Work}
\label{sec:related_work}
\subsection{Vulnerability Report Inconsistencies}
Mining software repositories has become a popular area of empirical software engineering research \cite{hassan2008road}. However, the use of these data sources does not come without perils; the data is not necessarily clean and can exhibit significant noise \cite{kalliamvakou2014promises, croft2021data}. Our study contributes to this body of knowledge by highlighting a unique data quality issue that is present in SV reporting data sources; SV severity ranking inconsistency. 

Tian et al. \cite{tian2016unreliability} identified that inconsistencies existed in bug severity data for duplicate bug reports, and hence demonstrated the unreliability of this data for bug severity prediction. Whilst duplicate bug reports can add inconsistency due to the differences from their respective reporters, we instead investigated the inconsistency of SV severity across data sources. 

NVD itself has acknowledged that it may exhibit inconsistencies in severity scores to other data sources\footnote{\url{https://nvd.nist.gov/general/FAQ-Sections/CVE-FAQs}}, due to subjective or inaccurate assessment. Several works have investigated inconsistency of CVSS scores for NVD in comparison to other vulnerability databases \cite{johnson2016can, jiang2020approach, schweigler2020investigation}, but they concluded that vulnerability databases were relatively consistent, due to the large overlap in the sources that they collected their data from. Hence, we investigated severity ranking inconsistencies from independent, rather than overlapping data sources. 

Finally, Dong et al. \cite{dong2019towards} investigated inconsistencies across NVD entries and their associated vulnerability reports, with a particular focus on vulnerability description and affected versions. They identified that information quality and consistency can be impacted as data is propagated from one source to another. We similarly investigated consistency of data propagation in the SV reporting lifecycle, but honed our focus on SV assessment and severity data.

\subsection{Automated Severity Prediction}
Several approaches have been proposed by researchers to automatically assign a severity level to a bug or security report \cite{gomes2019bug}, to increase automation and reduce developer efforts. One of the first notable contributions was by Menzies and Marcus \cite{menzies2008automated}, who proposed a rule learning technique to automatically assess severity of bug reports. Gomes et al. \cite{gomes2019bug} conducted a systematic mapping study of bug report severity prediction, and identified that the majority of works used learning-based approaches applied to unstructured text descriptions. 

Similarly, researchers have developed automated approaches to assess the severity of an SVs. Le et al. \cite{le2021survey} conducted a survey of automated software vulnerability assessment, and found that the majority of studies that aimed to predict severity used NVD text descriptions to predict CVSS scores. Recently, Le et al. \cite{le2021deepcva} proposed to automatically assess severity directly from source code, to reduce reporting delay. 

Hence, various data sources have been used by researchers to automatically assess severity at different stages; i.e., to increase automation for the initial reporting in bug reports, or for the more late and detailed analysis of NVD. To the best of our knowledge, we are the first to have compared and analysed inconsistencies in the data sources for this task.

\section{Conclusion}
\label{sec:conclusion}

In this study, we conducted a large scale analysis of SV severity reporting to identify characteristics, causes and impacts of inconsistency in SV severity data and reporting schemes, across the sources that it is reported. We performed this analysis using the Mozilla Firefox project as a case study. Predominantly, SVs were often ranked to be of lower severity at the time of reporting. This underestimation or lack of consideration for severity assessment caused such SVs to receive lower prioritization. We identified several potential causal factors for this initial inconsistent severity assessment, including that more experienced bug reporters often underestimated SV severity. We finally showed that this inconsistency in SV severity reporting schemes can severely degrade the performance of predictive tasks by up to 77\%, depending on the information source used. 

Through our findings we have drawn attention to this data problem and proposed several implications from this study. In summary, we recommend developers, researchers and security experts to deploy more consideration towards the quality and consistency of SV severity data sources. Furthermore, we have outlined some potential factors, and hence areas of improvement, that are correlated with correct assessment of SV severity at initial reporting. 

\section*{Acknowledgment}
This work has been supported by the Cyber Security Cooperative Research Centre Limited whose activities are partially funded by the Australian Government’s Cooperative Research Centre Programme.

\bibliographystyle{IEEEtran}
\bibliography{bibfile}

\end{document}